**The impact of hot charge carrier mobility on photocurrent losses in polymer-based solar cells**

*Bronson Philippa[1], Martin Stolterfoht[2], Paul L. Burn[2], Gytis Juška[3], Paul Meredith[2], Ronald D. White[1], and Almantas Pivrikas[2,*]*

[1] School of Engineering and Physical Sciences, James Cook University, Townsville 4811, Queensland Australia

[2] Centre for Organic Photonics & Electronics (COPE), School of Chemistry and Molecular Biosciences and School of Mathematics and Physics, The University of Queensland, Brisbane 4072, Queensland, Australia

[3] Department of Solid State Electronics Vilnius University 10222 Vilnius, Lithuania

* almantas.pivrikas@uq.edu.au

A typical signature of charge extraction in disordered organic systems is dispersive transport, which implies a distribution of charge carrier mobilities that negatively impact on device performance. Dispersive transport has been commonly understood to originate from a time-dependent mobility of hot charge carriers that reduces as excess energy is lost during relaxation in the density of states. In contrast, we show via photon energy, electric field and film thickness independence of carrier mobilities that the dispersive photocurrent in organic solar cells originates not from the loss of excess energy during hot carrier thermalization, but rather from the loss of carrier density to trap states during transport. Our results emphasize that further efforts should be directed to minimizing the density of trap states, rather than controlling energetic relaxation of hot carriers within the density of states.

While natural photosynthesis transfers electrons through a cascade of energy states, artificial photovoltaic systems must extract photogenerated charges to the electrodes, and despite recent performance gains[1], fundamental questions about this charge extraction still remain unanswered. There has been intense scrutiny of the mechanisms of charge generation and the impact of above-bandgap photon energy[2–4], however, this level of attention has not extended to studies of the *extraction* of such 'hot' charge carriers, despite the fact that efficient charge extraction is crucial for device performance[5].

The most characteristic feature of charge transport in disordered systems is the dispersion of the charge carrier movement velocities[6]. Dispersive transport harms device performance because the slowest carriers bring down the average mobility[7], and consequently, the vast majority of novel organic semiconductors remain inapplicable for efficient devices. Moreover, the detrimental effects of dispersion are exacerbated by the inhomogeneities in film thicknesses caused by the targeted low cost deposition methodologies, because the transit time distributions become dramatically longer and more dispersed in regions of increased thickness.

Dispersive transport in organic semiconductors is usually thought to be caused by the energetic relaxation of hot charge carriers within their density of states[8]. Spectroscopic measurements and Monte Carlo simulations have revealed energetic relaxation extending even to the microsecond timescales, where it could be relevant to bulk charge transport[9,10]. Even if the bulk of the energetic relaxation were to occur on very fast timescales, there is still the question of whether residual thermalization might continue to long, microsecond timescales. This energetic relaxation is often understood to cause a time-dependent mobility and therefore explain dispersive current transients[11,12], yet we will show here that this commonly-used model is inconsistent with our observations in high efficiency organic solar cell materials. Instead, there is an alternative mechanism for the creation of a distribution of carrier velocities, namely, via trapping. This observation has a very direct impact on the numerous models, theories and experimental results describing dispersive charge transport in disordered organic semiconductors. Furthermore, it points to a new strategy for improving charge transport "management" in devices such as organic solar cells.

The classic signature of dispersive transport is a time-of-flight photocurrent signal that decays with time even before the carriers have transited through the film[13]. This decay in photocurrent can occur due to two mechanisms, a reduction in carrier mobility, and/or a reduction in the number (or concentration) of moving carriers. The former, a time-dependent hot carrier mobility, is presently commonly believed to be the cause of dispersion in organic semiconductors[8,11,12] and it is usually understood to originate from a loss of energy as carriers thermalize within their density of states[14–16]. Higher energy carriers are expected to have a higher hopping probability, and hence a higher velocity[17–19], so the thermalization within the density of states causes the carrier mobility to decline. Recent studies [Melinana et al, Howard et al.] have reported mobility thermalization times on the order of microseconds. However, an alternative explanation for the decaying transient photocurrent, which is less commonly accepted in organic semiconductors, is a time dependent *concentration* that can arise if carriers are gradually lost to traps[20–22]. The photocurrent signal will continue to reduce as long as the net concentration of moving carriers continues to decrease. If that physical process prevails, there can be decaying photocurrent despite the moving carriers having a constant drift velocity. Additionally, if the cause of dispersion is trapping, then it will influence all devices, even those which operate in the dark[23,24].

In this article, we demonstrate that a time-dependent hot carrier mobility cannot explain the dispersive transport in our studied bulk heterojunction solar cells. We address this issue by performing transient photoconductivity experiments in which we vary the transit time by changing the electric field and/or device thickness. The expectation is that if the dominant cause of dispersive transport is mobility relaxation, then the average mobility and the amount of dispersion should vary with the electric field and/or film thickness, because longer transit times will allow for more relaxation to occur. Conversely, if the dominant effect is trapping, then it is the concentration of carriers which is changing in time rather than their mobility, and consequently, the average mobility and the dispersion range should not vary with film thickness or electric field. This transit time dependence allows these two dispersive mechanisms to be experimentally distinguished.

**Results**
**Numerical Simulations of Resistance dependent PhotoVoltage (RPV) measurements**
Our experiments were made possible by the development of a new transient photoconductivity technique that we call Resistance dependent PhotoVoltage (RPV), which is described here and in the Methods section. The experimental measurement circuit for RPV is shown in **Figure 1**. This setup is similar to time-of-flight, where charge carriers are photogenerated by a short low-intensity laser pulse. A low light intensity is necessary so that the electric field inside the device is undisturbed. The transient photosignal is determined by the competition between two simultaneous processes: the transport of charge carriers inside the film, and the response of the external RC circuit. Unique to the RPV approach, and in contrast with time-of-flight, the entire measurement is repeated at many different load resistances spanning the range from differential mode (small $R$) to integral mode (large $R$).

The resistance is varied for two reasons: firstly, to visualize the transit times, as will be shown below; and secondly, to reveal the slower carrier mobility by amplifying the slower carrier's conduction current. The slower carriers produce a much smaller current than the faster carriers, and their transit would be buried in the noise at resistances that are optimized for the faster carriers. Slower carriers have much longer transit times, allowing the use of larger resistances, and consequently allowing for their weaker electrical signal to be amplified. In this way, RPV bridges the gap between differential mode and integral mode time-of-flight, and allows measurement of the transport of both types of charge carriers.

The combination of the *RC* circuit dynamics, dispersive transport, and optical interference effects prevent analytic analysis of the transients. To study highly dispersive systems, such as organic solar cells, the simultaneous impact of all these effects must be understood. We applied numerical simulations to develop this understanding. The simulations are described in the Methods section and in the Supplementary Information. Typical simulated transients for an organic solar cell with dispersive transport are shown in **Figure 2 (a)**. The transients show two distinct extraction "shoulders," as indicated by the arrows. The transients at different resistances assist in visually identifying the location of these "shoulders." The positions of these arrows correspond to the mean transit times required for the faster and slower carriers to cross the entire thickness of the film. In this simulation, carriers are repeatedly trapped and de-trapped, creating dispersion because the total time spent in traps is different for different carriers. The resulting distribution of transit times is shown at the top of Figure 2 (a), and its approximate width is indicated by the shaded background. It can be seen that the RPV technique allows the mean charge carrier mobility to be obtained even in the presence of strong dispersion.

In addition to shallow traps that cause dispersion, we also considered deep traps that immobilize carriers for times much longer than the transit time of either carrier. Long lived trapping is typical in disordered organic semiconductors[22,25], because many organic materials behave as unipolar conductors, and solar cells often have strongly imbalanced mobilities[26]. In these cases, repeated photogeneration adds more trapped charge in the form of the immobilized charge carriers, which might accumulate with every repetitive laser shot, redistributing the electric field and distorting the measurement. **Figure 2 (b)** shows simulations of this film charging for the case of fast Langevin-type recombination under repeated laser shots, as would arise from the presence of deep trap states far inside the forbidden energy gap[27]. These are large resistance transients, in other words, the measurement circuit has integrated the photocurrent such that the peak voltage is proportional to the extracted charge. If the extracted charge is decreasing and the extraction time remains constant, then carriers must be lost to recombination and not due to field screening, and hence we conclude that the trapped charges act as recombination sites for the mobile carriers. However, the mobility of the charge carriers can be determined independently of the trapping effects, because the rapid Langevin recombination prevents the build-up of large amounts of trapped charge that would disturb the transit time.

**Experimental measurements**

We chose to study the well-known photovoltaic blend[28] of poly[*N*-9''-hepta-decanyl-2,7-carbazole-*alt*-5,5-(4',7'-di-2-thienyl-2',1',3'-benzothiadiazole)] (PCDTBT) and [6,6]-phenyl-$C_{70}$-butyric acid methyl ester ($PC_{71}BM$) in an optimized blend ratio of 1:4 by weight.[29] This blend is ideally suited to this study because its amorphous nature allows the elimination of any film thickness dependent morphology[30]. In order to see the generality of observed effects, we have also done the same experiments on poly[[4,8-bis[(2-ethylhexyl)oxy]benzo[1,2-b:4,5-

b']dithiophene-2,6-diyl] [3-fluoro-2-[(2-ethylhexyl)carbonyl] thieno[3,4-b]thiophenediyl]] (PTB7):PC$_{71}$BM blends, the results of which are shown in the Supplementary Information. The thin film (active layer thickness of 75 nm) PCDTBT:PC$_{71}$BM solar cell exhibited a power conversion efficiency of 6.3% under standard AM1.5G illumination, while the PTB7:PC$_{71}$BM blends reached 7.7%. Current-voltage curves for both devices are shown in Supplementary Figure 5. None of the optimized PCDTBT or PTB7 based devices demonstrated any significant film morphology inconsistencies in the range of studied film thickness. (See the Methods section for the details of the fabrication and the Supplementary Information for characterization of photovoltaic performance). The presence of dispersive transport was confirmed by time-of-flight experiments on thick films (Supplementary Figure 6). No photocurrent plateaus were observed; the transients decrease with time as is typical of dispersive systems.

**Figure 3** shows the recorded RPV transient signals for a PCDTBT:PC$_{71}$BM solar cell. All transients were recorded at near to short-circuit conditions. This remains true even at large resistances, because the maximum photovoltage occurring during the transient is substantially less than the built-in voltage. The first shoulder marks the arrival time of faster carriers (27 ns), which is attributed to electron transport since the time scale is similar to that measured for PC$_{71}$BM (please refer to the Supplementary Information for measurements on PC$_{71}$BM). The second shoulder is less well defined due to the strongly dispersive nature of hole transport in this system, but marks the arrival of the slower carriers (2.59 µs). Mean electron and hole mobilities were determined from the shoulders in the transients, as indicated by arrows in Figure 3, with the approximate spread of arrival times indicated by the shaded boxes (corresponding to the regions where the transients deviate from the dotted lines). The edges of these shaded boxes give the "fastest" and "slowest" case transit times, from which we obtained the dispersion range in the mobilities for each species. This range is an essential feature of the dispersive transport exhibited by this system, because a single mobility value does not correctly quantify the transport when the system is dispersive. We measured the mean electron mobility to be $2.9 \times 10^{-3}$ cm$^2$ V$^{-1}$ s$^{-1}$ with a dispersion range from $1.1 \times 10^{-3}$ cm$^2$ V$^{-1}$ s$^{-1}$ to $4.5 \times 10^{-3}$ cm$^2$ V$^{-1}$ s$^{-1}$ and the mean hole mobility to be $3 \times 10^{-5}$ cm$^2$ V$^{-1}$ s$^{-1}$ with a dispersion range from $9.2 \times 10^{-6}$ cm$^2$ V$^{-1}$ s$^{-1}$ to $7.4 \times 10^{-5}$ cm$^2$ V$^{-1}$ s$^{-1}$. Despite the high level of dispersion observed here (the hole dispersion range covers nearly an order of magnitude), the OPV device still maintains good performance. However, further work is necessary to identify the impact of the dispersion range on the performance of solar cells.

Next, we studied the impact of photon energy on the hot charge carrier transport, because any relaxation effects are likely to be dependent upon the initial energy. This is important because of recent suggestions that excess above-bandgap energy may assist excitonic dissociation[4], although the methodology of that observation has been challenged[31]. We note that quantum yields have been shown to be independent of the energy level of the excited state, suggesting that hot excitons are indeed not beneficial for exciton separation[32]. Nevertheless, hot *charge carriers* – rather than excitons – might also possess excess energy and shape the internal quantum efficiency spectra; therefore, it is important to clarify these effects, aiming for improvement in the charge extraction of typical low mobility organic materials. In the past the absence of hot charge carrier effects has been observed indirectly[33]. Numerical simulations predict that RPV is independent of optical interference effects (Supplementary Figures 2 and 3), allowing direct and unambiguous measurement of any hot charge carrier effects that may be present. RPV transients were measured at two different photon energies, 3.49 eV (355 nm) and 2.33 eV (532 nm). The results are plotted in **Figure 4**, showing nearly identical transients resulting from laser excitation at the two different wavelengths. The photon energy independent mobility suggests that excess energy plays a minimal role in dispersive transport, since carrier thermalization (if it is present) must happen in time scales much shorter than the transit time.

To further confirm that the dispersion in hot carrier mobilities is not caused by the thermalization of carriers, we studied the electric field and film thickness dependence. Longer transit times should allow more time for thermalization, thus influencing the result if the dispersion is due to carrier relaxation. The results are shown in **Figure 5**; the Supplementary Information includes a selection of the transients from which these mobilities were estimated. The mobilities and dispersion ranges are completely independent of electric field and photon energy [Figure 5 (a)], suggesting that trapping mechanisms are more significant than relaxation mechanisms. The lack of electric field dependence is in contrast with the Poole-Frenkel dependence reported in pristine PCDTBT[34]. This is an unexpected result, because in disordered organic systems significant electric field dependence is typically observed, even at relatively low values of electric fields[34], which is thought to originate from hopping-type charge transport. Further studies of the temperature dependence, and measurements on other systems, have to be performed in order to clarify the origin of this observation. Additionally, we observe that the mean mobilities and dispersion ranges are nearly independent of the film thickness [Figure 5 (b)]. We attribute the small changes in mobility to device-to-device variations that result from the fabrication process. The thickness independence of the mean mobilities and dispersion ranges further support the claim that the dispersion is caused by traps instead of relaxation. A charge carrier density dependence in the mobility even at low concentrations has been observed in P3HT:PCBM blends[24], and we note that a concentration dependence might cause dispersion as carriers gradually become trapped and the density decreases. We do not exclude the possibility of a density dependence here. However, in our measurements, increasing thickness corresponds to lower densities because the amount of photogenerated charge was always less than $CU$, which is inversely proportional to thickness. Consequently, the thickness independence in the mobility implies that there is negligible density dependence at the concentrations probed here.

Further measurements were also performed on solar cells made with PTB7 blends. The results show the same conclusions as the PCDTBT blends: the mean mobility and dispersion ranges are independent of film thickness, applied electric field, and photon energy (Supplementary Figures 8, 10, and 11). The results reported here appear to be generally applicable and are certainly not specific to PCDTBT blends.

**Discussion**

Charge transport in the studied operational OPV blends is strongly dispersive, as demonstrated by the decaying time-of-flight photocurrent transients in thick devices (Supplementary Figure 6). These time-of-flight transients were recorded in a regime where drift dominates over diffusion, so the current density is described by $j = eE\,(n\mu_n + p\mu_n)$, where $e$ is the charge of an electron, $\mu_n$ and $\mu_p$ are the electron and hole mobilities, $n$ and $p$ are the carrier concentrations, and $E$ is the electric field. The observation of a decaying photocurrent density $j$ can be explained by two mechanisms: thermalization (a time dependent mobility, $\mu$), and/or trapping (a time dependent concentration of moving charge carriers, $n$). These mechanisms are schematically illustrated in Figure 6, from which it can be seen that either model would result in dispersive photocurrent transients. We found no evidence of thermalization-type effects on the timescales comparable with those involved in charge transport. Figure 4 directly demonstrates that that excess energy of hot carriers has essentially no contribution to mobility or dispersion. In Figure 5, we demonstrate that the dispersion range is independent of the applied electric field and changes very little with thickness. If thermalization on transport time scales[16] were the cause of the dispersion, then modifications to the transit time should change the mean mobility and/or dispersion range by varying the time available for relaxation. Such a variation was not observed, and hence we exclude thermalization as the mechanism of the dispersive transport. Any relaxation processes must be much faster than charge transport, so that the distance covered by charges as they relax is insignificant compared with the film thickness, and hence the relaxation has negligible contribution to the overall dispersion. With relaxation excluded, the only remaining mechanism is a reduction in the concentration of moving carriers, therefore, we conclude that *trapping* is the primary cause of the dispersion in these systems. This challenges the widely-used model of hot carrier relaxation within the density of states. Consequently, dispersive transport potentially impacts on the many different devices that employ films made from disordered semiconductors, including those that operate in the dark or at steady-state conditions.

In conclusion, electron and hole mobilities and their dispersion ranges were measured simultaneously using the RPV technique in a high efficiency narrow optical gap polymer/fullerene system (PCDTBT:$PC_{71}BM$). We found that the transport of electrons and holes are both strongly dispersive in these thin, efficient solar cells. We introduced the dispersion range as a parameter to quantify charge transport, since a single mobility value is insufficient to properly characterize a dispersive material. We directly observed the absence of "hot carrier" effects on time scales relevant to charge extraction, and furthermore found that the dispersion is caused by trapping rather than thermal relaxation. We have found that the widely-used model of hot carrier relaxation within a density of states is not the dominant process causing the dispersion in the studied solar cells. Furthermore, in contrast with the Poole-Frenkel dependence previously reported in pristine PCDTBT and other disordered systems, the studied solar cell blends exhibit an unexpected negligible electric field dependence. While further work is needed to clarify this observation, electric field independence may assist in maintaining a good fill factor by keeping the mobility higher near the maximum power point. The absence of hot carrier effects and an electric field independent mobility were also observed in PTB7:PC$_{71}$BM solar cells, suggesting that these conclusions may be more generally applicable. This work signifies the importance of localized trap states as opposed to thermalization and hot carrier effects in efficient polymer-based solar cells. Since dispersion arises from trapping, it is also important for other types of devices, such as organic field effect transistors and diodes. Trap states are relevant whether the carriers were injected or photogenerated, and whether the device is in transient or equilibrium conditions. Our results suggest that further scientific research should be directed towards reducing the density of trap states rather that utilizing above-bandgap energy for improving electronic device performance.

**Methods**
*Numerical simulations:* The simulations are based on a standard one-dimensional drift-diffusion-recombination solver[35,36] assuming a negligible amount of equilibrium carriers, which is typically the case in organic semiconductors[37] as well as in the studied devices. For simulations of dispersive transport, we implemented a multiple trapping and release model[20,21,38] with an exponential density of localized states. The full list of equations are given in the Supplementary Information.

*Solar cell fabrication:* 15 Ω/sq. Indium tin oxide (80 nm thick, purchased from Kintec) coated glass substrates were cleaned in a 100 ºC water bath with alconox (detergent), followed by sonicating in sequence with de-ionized water, acetone and 2-propanol for 6 minutes each. Next, a 30 nm layer of poly(3,4-ethylenedioxythiophene):poly(styrenesulfonate) (PEDOT:PSS) was spin-coated at 5000 rpm for 60 sec onto the cleaned substrates, which were then annealed at 170 ºC for a few minutes in air. For PCDTBT devices, a solution of PCDTBT (purchased from SJPC Group) and $PC_{71}BM$ (purchased from Nano-C) was prepared by using a 1:4 blend ratio by weight and a total concentration of 25 mg/cm$^3$ in dichlorobenzene (DCB). Solar cells with four active layer thicknesses, 75 nm, 230 nm, 270 nm and 410 nm (measured by a DekTek profilometer), were fabricated by spin coating. For PTB7 devices, the active layer of PTB7 (1-Material, Mw = 97.5 kDa, PDI = 2.1) and $PC_{71}BM$ (ADS) was prepared as previously described[1] resulting in 100 nm, 150 nm, 230 nm, and 700 nm thick films. To complete the solar cells 1.2 nm of samarium and 75 nm of aluminium were deposited under a $10^{-6}$ mbar vacuum by thermal evaporation. The device areas were 0.2 cm$^2$ for current density versus voltage (*J-V*) measurements and 3.5 mm$^2$ for charge transport measurements. *J-V* characteristics were obtained in a 4-wire source sense configuration and an illumination mask was used to prevent photocurrent collection from outside of the active area. An Abet solar simulator was used as the illumination source and provided ~ 100 mW/cm$^2$ AM1.5G light.

*RPV measurements:* A delay/trigger generator (Stanford Research Systems DG535) was used to trigger the laser and function generator (Agilent 33250A) pulses for timing control. A pulsed Nd:Yag laser (Brio Quantel) with a pulse length of 5 ns was used to generate the carriers. Optical filters were used to reduce the laser intensity for the RPV measurements. A function generator was used to apply external voltage pulses for electric field dependent mobility measurements. RPV photovoltage signals were recorded with an oscilloscope (WaveRunner 6200A) at various external load resistances. RPV transients were smoothed with an adjacent averaging function to neutralize the electromagnetic wave oscillations in the measurement circuit. In agreement with previous studies done by Clarke *et al.*,[39] dark-CELIV transient responses showed no equilibrium carrier extraction, justifying the application of RPV to the studied devices. Optical interference simulations were performed using the transfer matrix approach[40] with typical optical constants of PCDTBT/PCBM blends.[41]

**Acknowledgements**
We thank Jao van de Lagemaat for providing the software for transfer matrix optical interference simulations, and Ardalan Armin for assistance with the optical simulations. Computational resources were provided by the James Cook University High Performance Computing Centre. AP is the recipient of an Australian Research Council Discovery Early Career Researcher Award (Projects: ARC DECRA DE120102271, UQ ECR59-2011002311 and UQ NSRSF-2011002734). PM and PLB are a Vice Chancellor's Senior Research Fellows. Research at Vilnius University was funded by the European Social Fund under the Global Grant measure. We acknowledge funding from the University of Queensland (Strategic Initiative – Centre for Organic Photonics & Electronics), the Queensland Government (National and International Research Alliances Program). This work was performed in part at the Queensland node of the Australian National Fabrication Facility (ANFF) - a company established under the National Collaborative Research Infrastructure Strategy to provide nano and microfabrication facilities for Australia's researchers.

**Author contributions**
B.P. and A. P. wrote the manuscript and interpreted the results. B. P. wrote the software and performed the simulations. M. S. fabricated the devices and performed the measurements. P. L. B. and P. M. supervised the experimental study. R. D. W. supervised the theoretical study and assisted with developing the software and simulations. G. J. contributed to the experimental design. A. P. conceptualized the experiment.

**Additional Information**

The author(s) declare no competing financial interests.

**Figure captions**

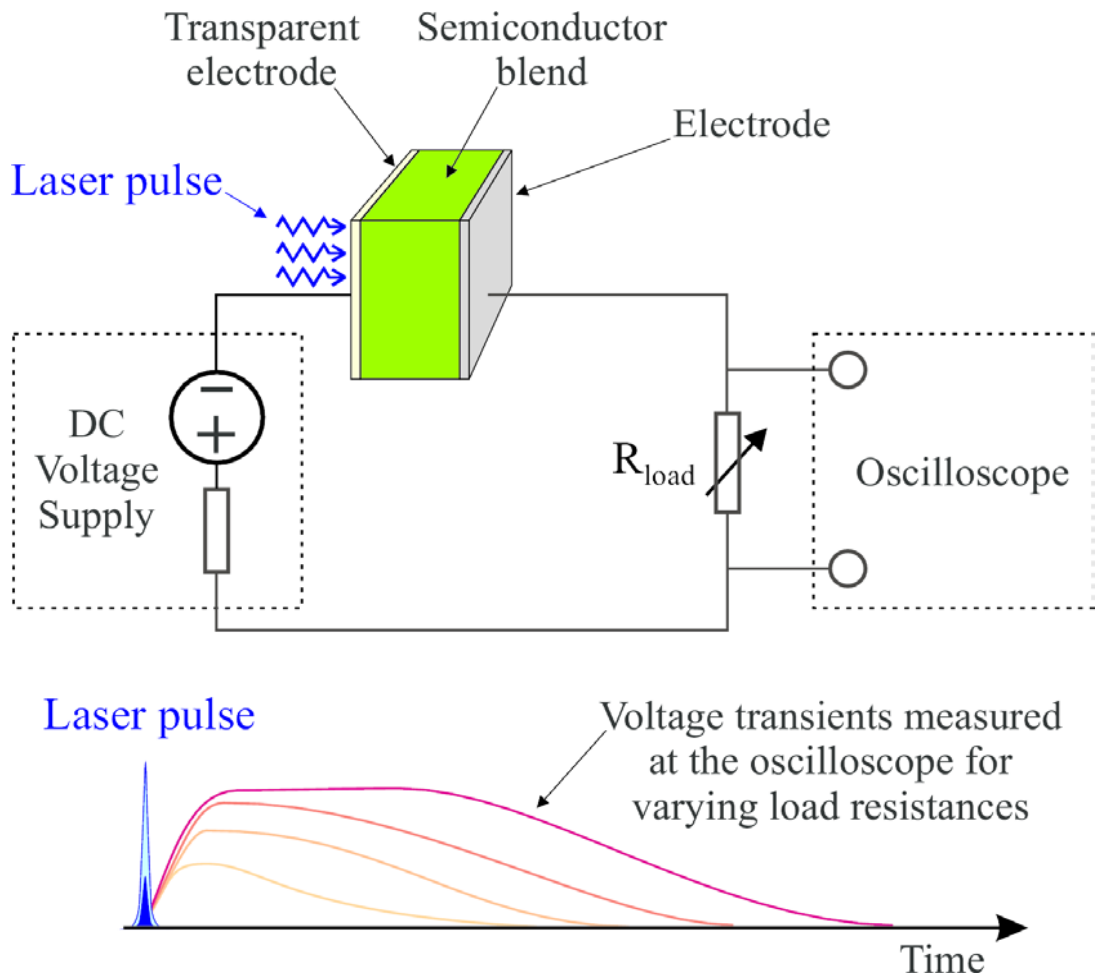

**Figure 1.** Resistance dependent PhotoVoltage (RPV) measurement circuit (**top**) and timing diagram (**bottom**). A low light intensity nanosecond laser pulse is used to photogenerate charge carriers inside (for example) the semiconductor junction of an organic solar cell. Low light intensity is critical in the RPV experiment to ensure operation within the "small charge extraction mode" where the internal electric field distribution in the film is not altered by transported charges. After photogeneration, the charge carrier transport through the film is driven by the built-in or the applied external electric field, and the resulting transient photosignal is recorded by an oscilloscope. The transient photosignals are measured at various load resistances $R_{load}$.

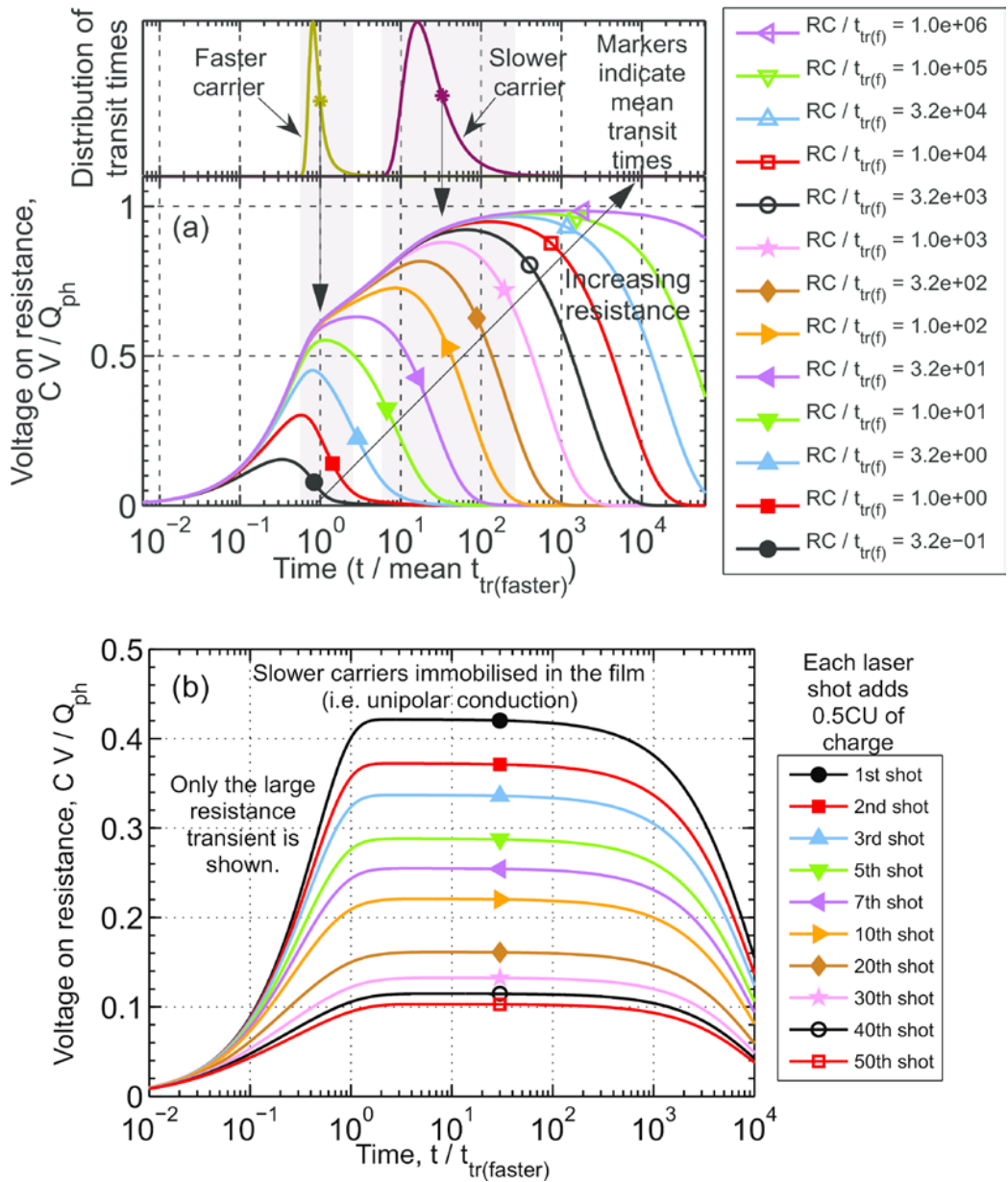

**Figure 2.** Numerically simulated RPV transients in the case of **(a)** dispersive transport caused by shallow traps, and **(b)** film charging caused by deep traps. **(a)** In the case of dispersive transport, the extraction "shoulders" approximately correspond to the mean charge carrier mobility. **(b)** In the case of deep traps, the film becomes charged and the magnitude of the RPV transient is reduced in subsequent shots of the laser, but the transit "shoulder" remains unhindered which allows for reliable charge carrier mobility estimation.

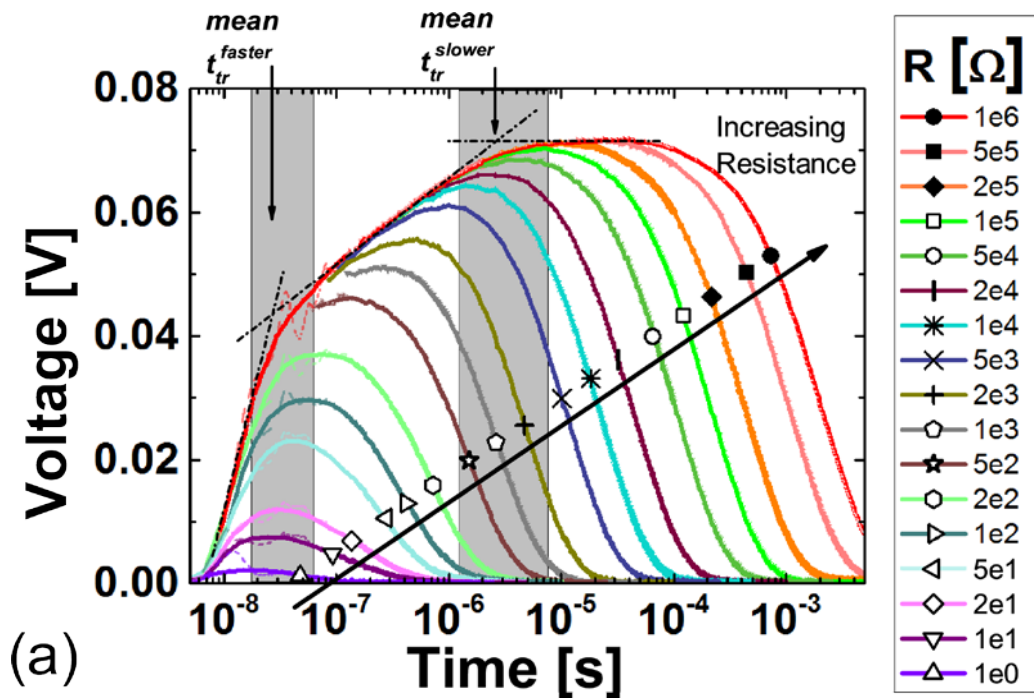

**Figure 3**. Experimentally measured RPV transient photo-signals in an optimized PCDTBT:PC$_{71}$BM solar cell. Mean electron (faster) and hole (slower) transit times are marked, from which the respective mean mobilities are estimated. The dispersive nature of charge transport in the studied solar cells is highlighted by shaded boxes marking the range of carrier arrival times. Thin curves show recorded data, while bold lines show data smoothed by adjacent averaging. The short timescales for large resistances were omitted for clarity.

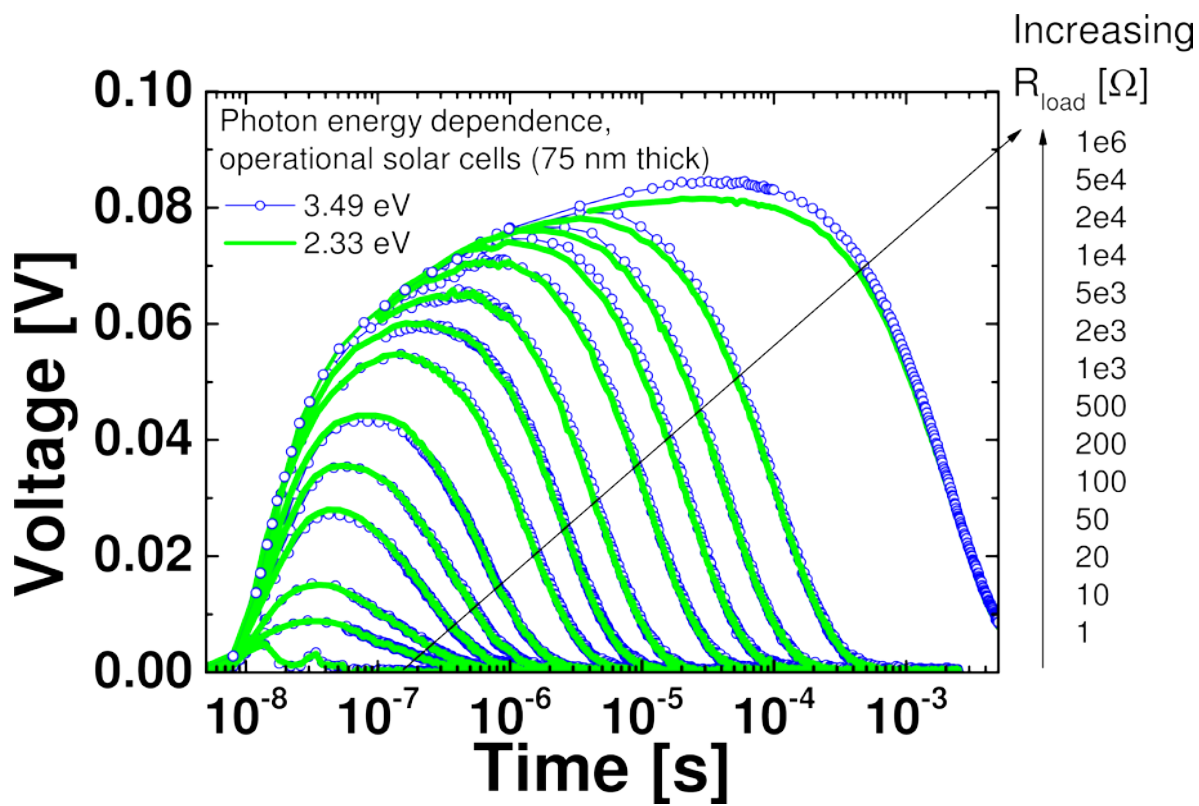

**Figure 4.** RPV transients measured on a 75 nm PCDTBT:PC$_{71}$BM solar cell using two different laser wavelengths: 355 nm (3.49 eV) and 532 nm (2.33 eV). The nearly identical transient responses directly demonstrate the absence of hot carrier effects in this system.

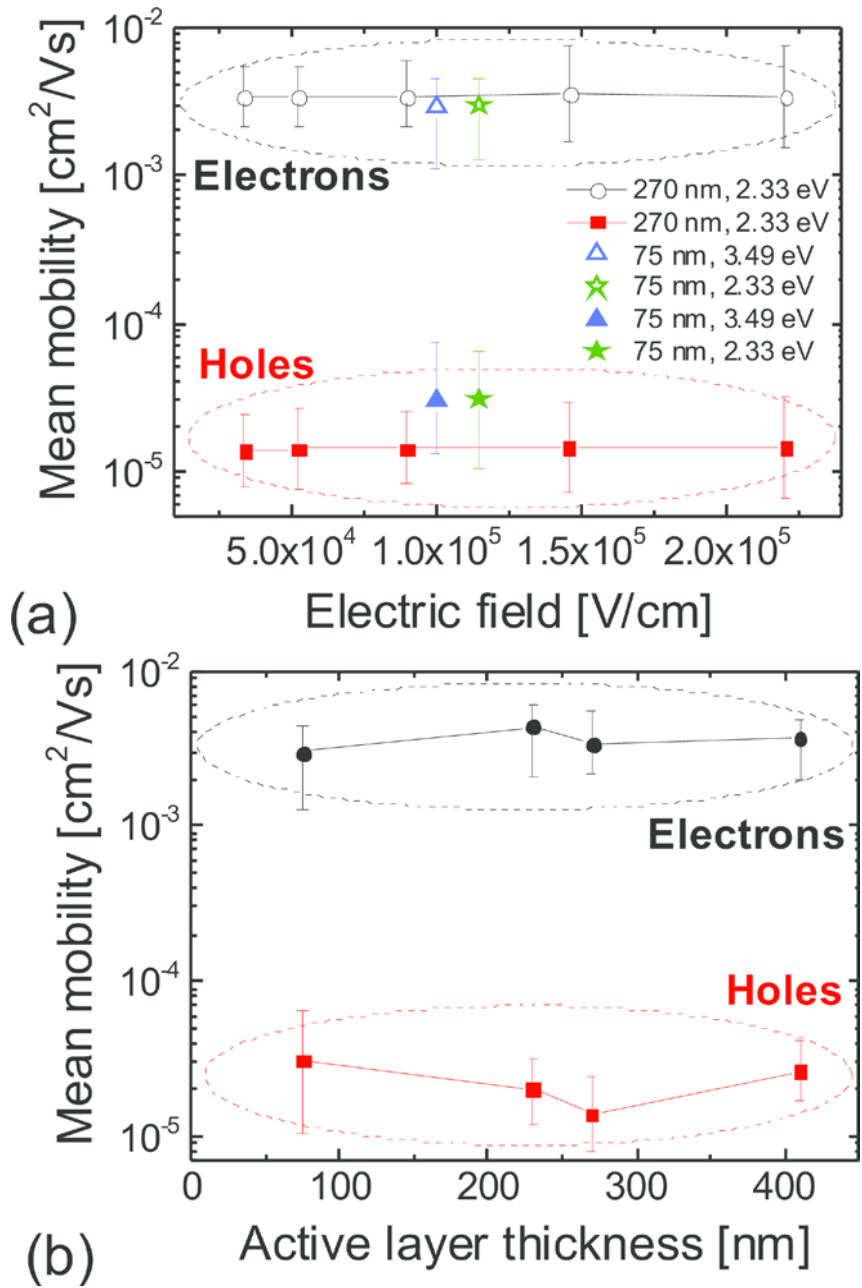

**Figure 5.** Electron and hole mobilities measured in PCDTBT:PC$_{71}$BM solar cells. The error bars show the dispersion ranges. Carrier mobilities and dispersion ranges are independent of electric field and photon energy [panel (**a**)], and nearly independent of film thickness [panel (**b**)], demonstrating that carrier thermalization cannot account for the dispersive transport in this system. Consequently, dispersion is caused by trapping.

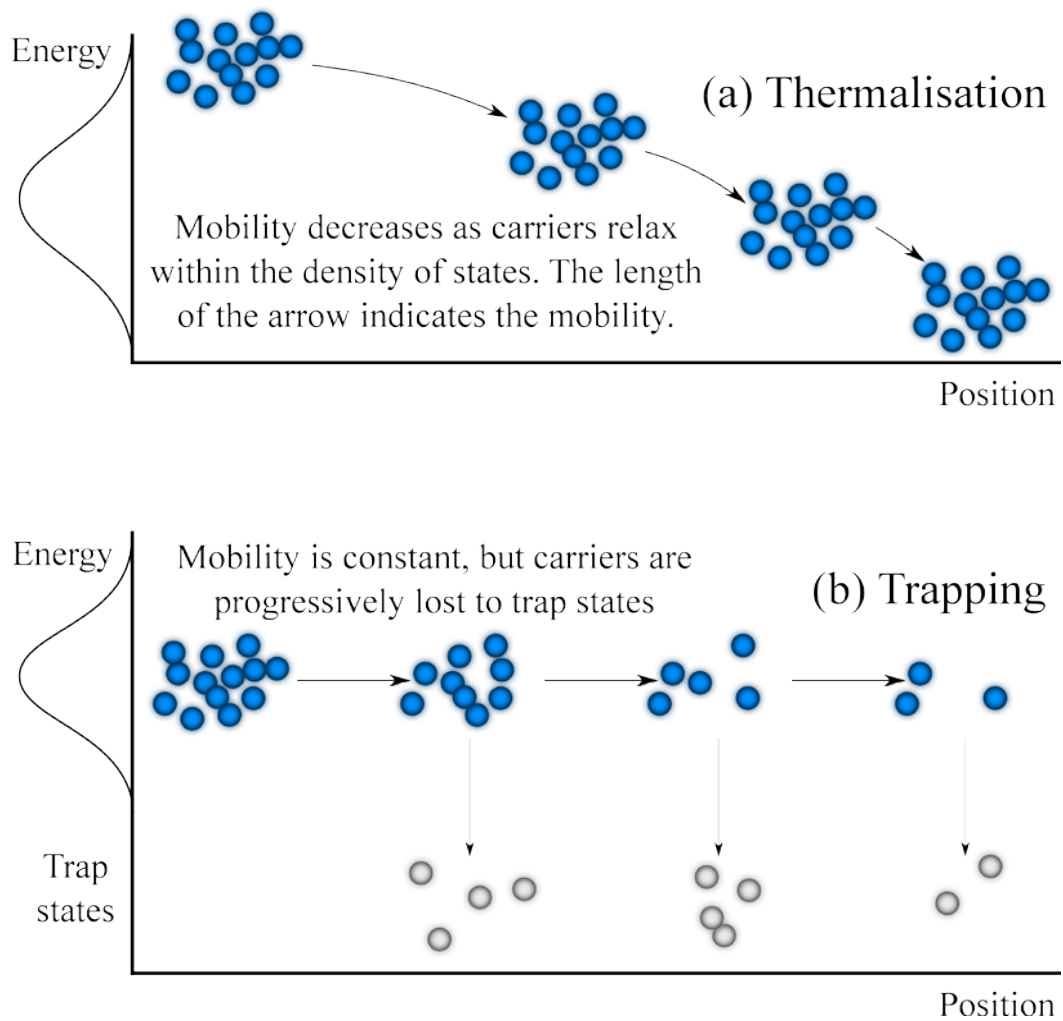

**Figure 6.** Schematic illustration of the two pathways to dispersive transport. **(a)** Thermalisation causes the mobility to decrease with time, whereas **(b)** trapping causes the loss of carrier density. We have shown here that the latter case (trapping) is the dominant effect in the studied solar cells.

# Supplementary Information for:

# The impact of hot charge carrier mobility on photocurrent losses in polymer-based solar cells

*Bronson Philippa[1], Martin Stolterfoht[2], Paul L. Burn[2], Gytis Juška[3], Paul Meredith[2], Ronald D. White[1], and Almantas Pivrikas[2,\*]*

[1] School of Engineering and Physical Sciences, James Cook University, Townsville 4811, Queensland Australia

[2] Centre for Organic Photonics & Electronics (COPE), School of Chemistry and Molecular Biosciences and School of Mathematics and Physics, The University of Queensland, Brisbane 4072, Queensland, Australia

[3] Department of Solid State Electronics Vilnius University 10222 Vilnius, Lithuania

\* almantas.pivrikas@uq.edu.au

**RPV experiment**

Volume photogeneration, as will occur in thin films, results in a spread of arrival times as different charge carriers travel different distances. Gradually increasing the load resistance assists in revealing these arrival times by lengthening the time scale of the measurement to incorporate those carriers that travel further. With a large enough resistance, the peak voltage location saturates, indicating complete extraction of the corresponding type of carrier, as can be seen below in Supplementary Figure 1. In this way, the final saturated peak location reveals the transit time $t_{tr}$ of those carriers that transited the entire film. Saturation of the peak voltage is thus an important indicator of complete carrier extraction, and failure to observe this saturation could result in an underestimation of the transit time. The mobility $\mu$ is then calculated from the transit time using the equation $t_{tr} = d^2\mu^{-1}U^{-1}$, where $d$ is the film thickness and $U$ is the sum of the built-in voltage and the applied voltage. Experimentally, the amount of photogenerated charge ($Q_{ph}$) needs to be kept small, such that $Q_{ph} < CU$, in order to avoid the space charge effects that would redistribute the electric field and disturb the transit time. This condition is easily checked experimentally by confirming that the maximum photovoltage is at least 10 times smaller than $U$.

**Non-dispersive (ideal case) simulations**

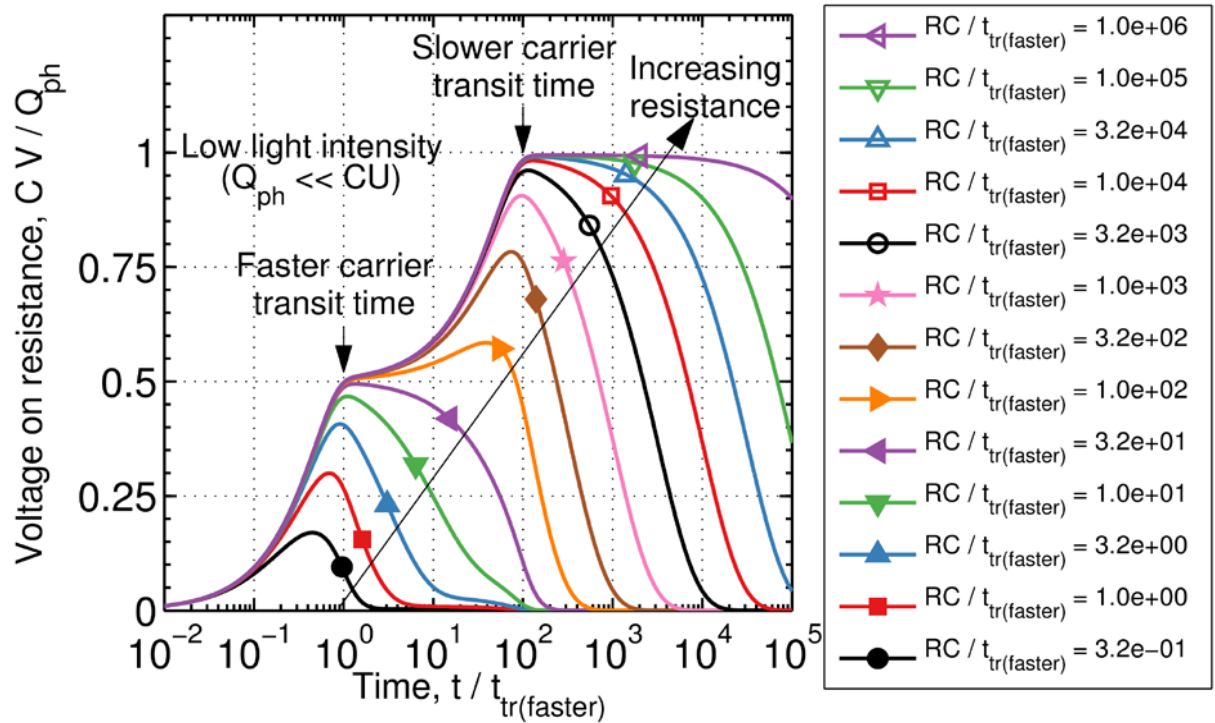

**Supplementary Figure 1.** Simulated RPV transients in the ideal case with non-dispersive transport. The transit times of both faster and slower carriers are clearly visible by the extraction shoulders.

**Light absorption independence**

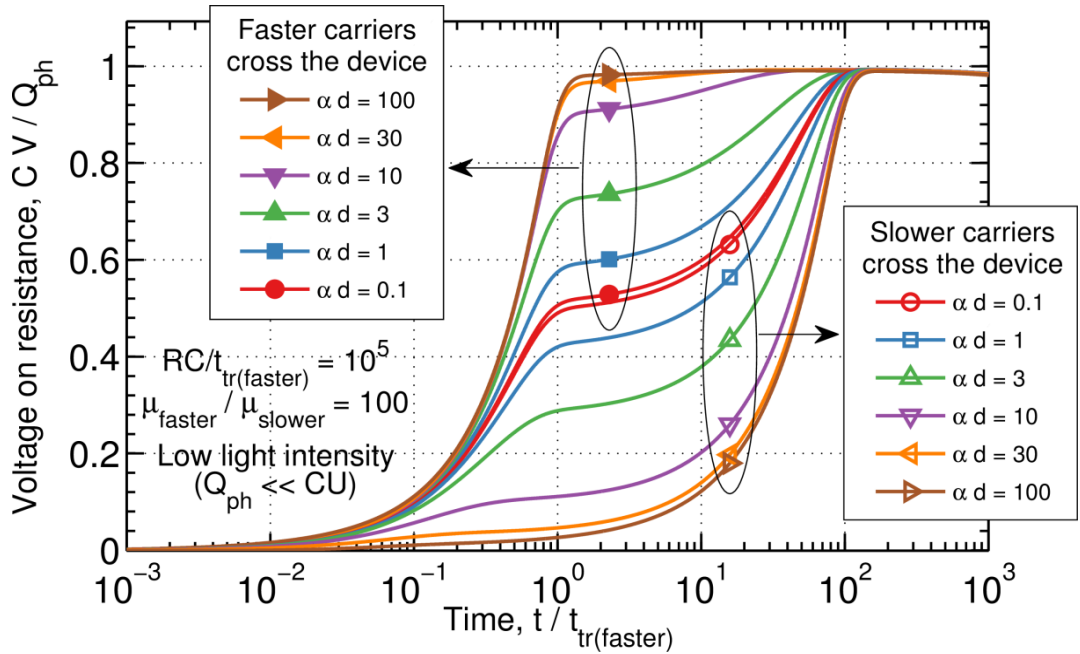

**Supplementary Figure 2.** Numerically calculated transients at high load resistance, varying the light absorption profile. The peak locations marking the carrier transit times are largely independent of the absorption profile, allowing reliable estimation of transit times. The light absorption profiles were calculated using the Beer-Lambert law, i.e. under the condition that the photogenerated charge distribution is proportional to $e^{-\alpha x}$, where $\alpha$ is the absorption coefficient and $x$ is the spatial coordinate normal to the electrodes. These profiles are characterized by the dimensionless product $\alpha d$. In the case of volume generation ($\alpha d \leq 3$), transit time information can be obtained for both carriers, but for surface generation ($\alpha d \geq 10$), it is only possible to observe the transit time for the carrier that actually transits through the film.

**Light interference independence**

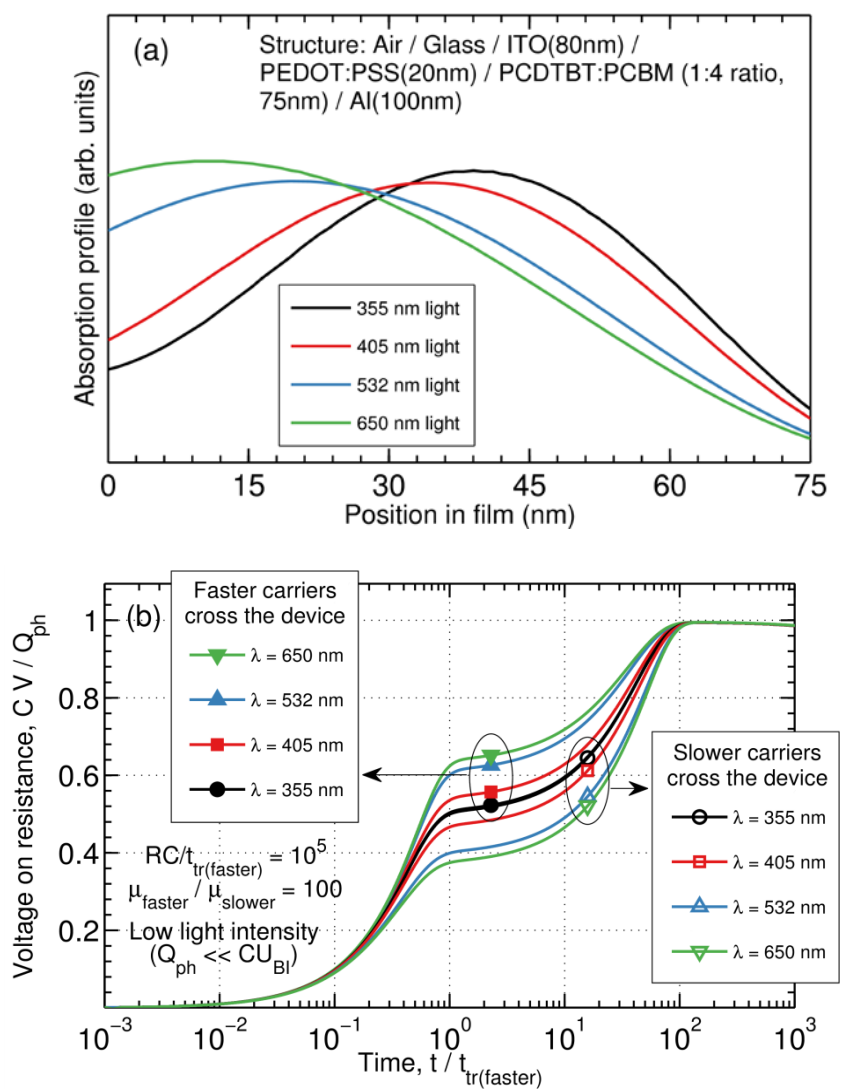

**Supplementary Figure 3.** **(a)** Optical interference (cavity) governed charge carrier photogeneration profiles in the active layer of the 75 nm thick organic solar cell shown for different excitation wavelengths. Note how the photocarrier extraction distance varies with wavelength, which must be accounted for when calculating charge carrier mobilities from transit time. **(b)** RPV transients at high load resistance simulated for the corresponding photocarrier generation profiles shown in (a). While the heights of the shoulders vary, their positions are independent of the absorption profile, demonstrating the negligible optical interference effects for carrier mobility calculations.

**RPV transients for the case of balanced mobilities**

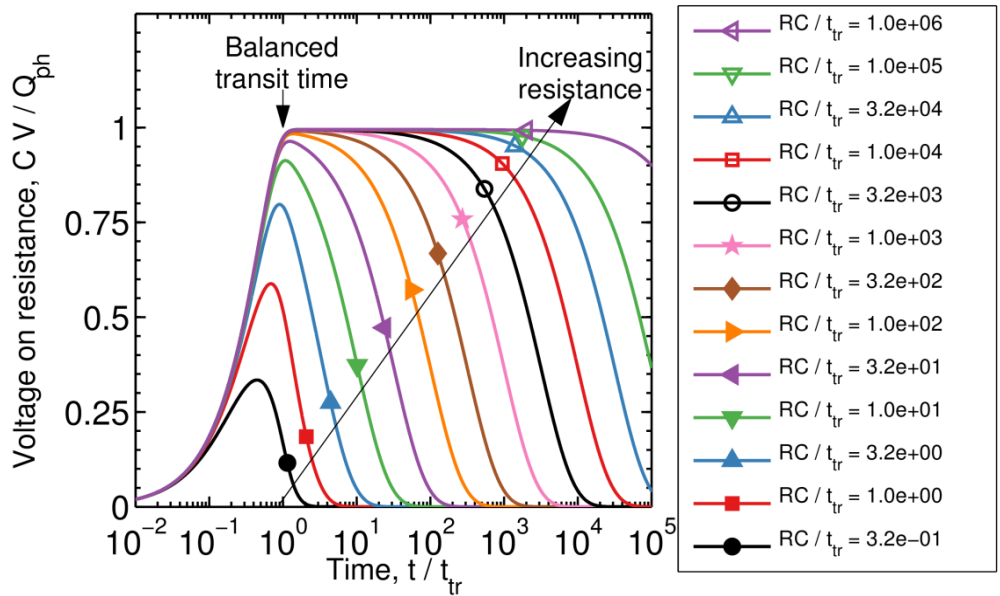

**Supplementary Figure 4.** Calculated transients in the case of balanced mobilities. Only a single transit time is visible, as indicated by the arrow.

**PCDTBT/PC$_{71}$BM and PTB7:PC$_{71}$BM photovoltaic performance**

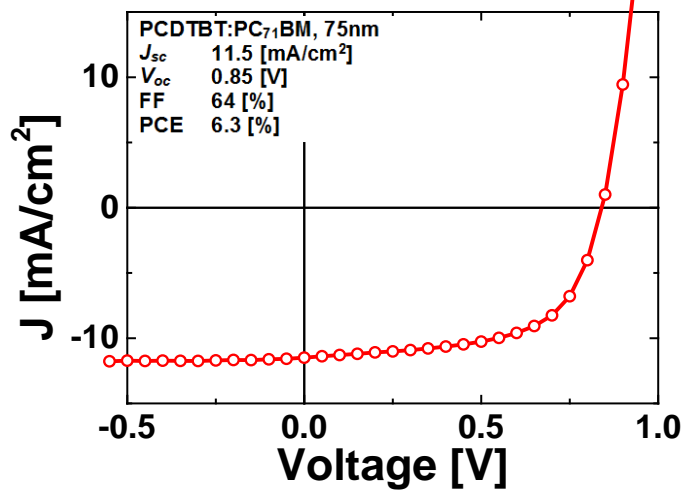

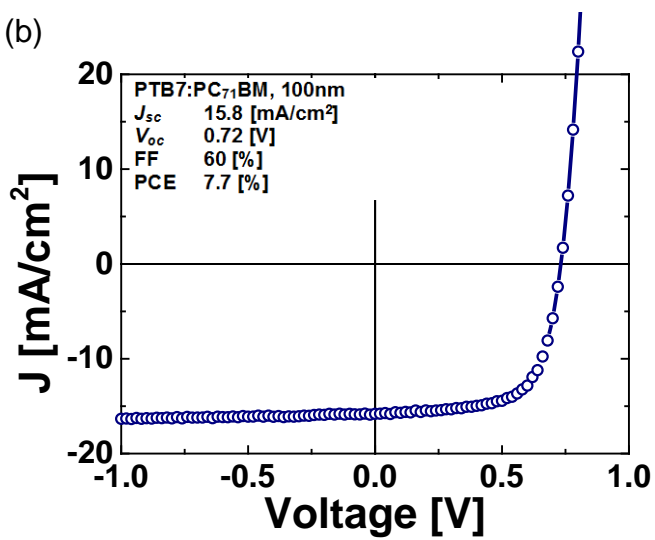

**Supplementary Figure 5.** J-V characteristic of typical (a) PCDTBT:PC71BM and (b) PTB7:PC71BM solar cells under AM 1.5G illumination demonstrating state-of-the-art efficiencies with these device combinations in optimized blends of (a) 1:4 and (b) 1:1.5 polymer to fullerene by weight.

**Dispersive time of flight transients**

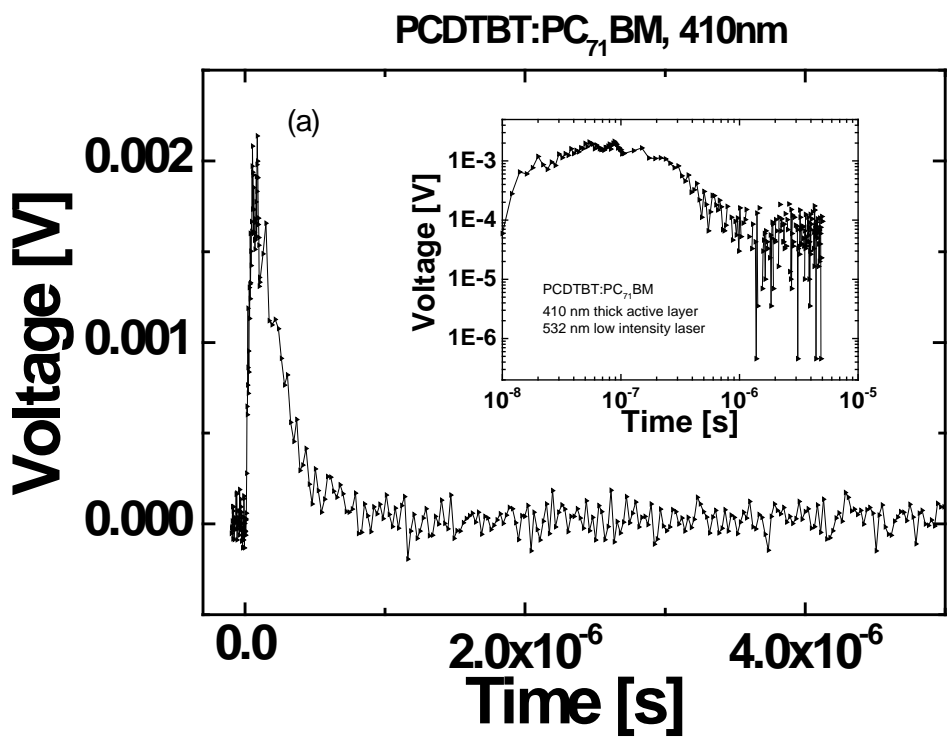

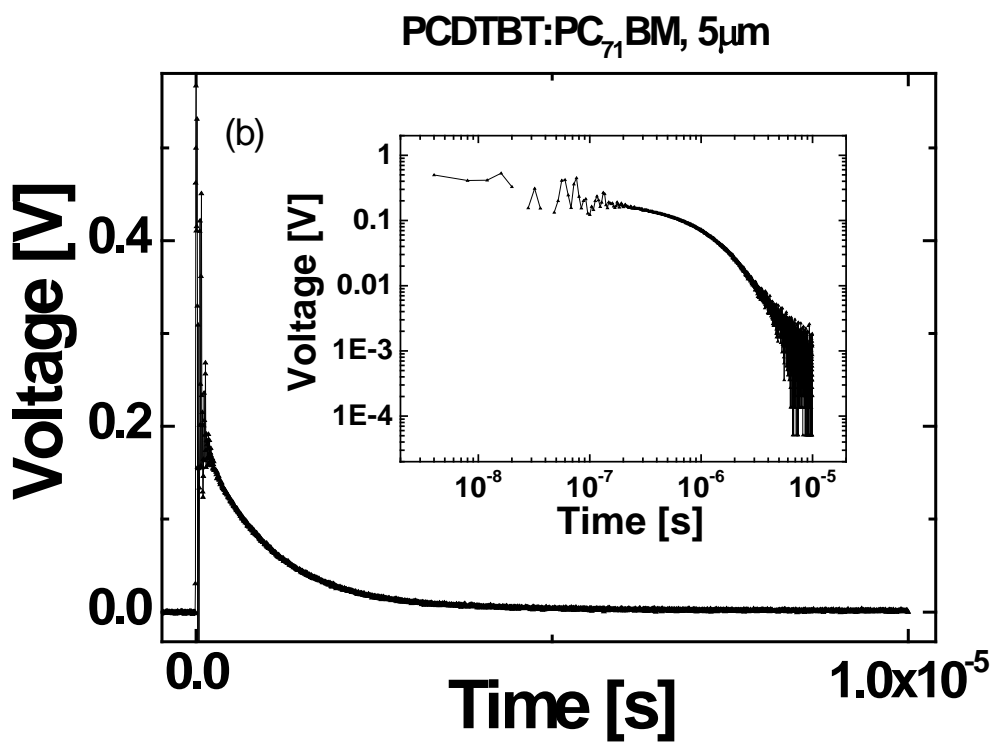

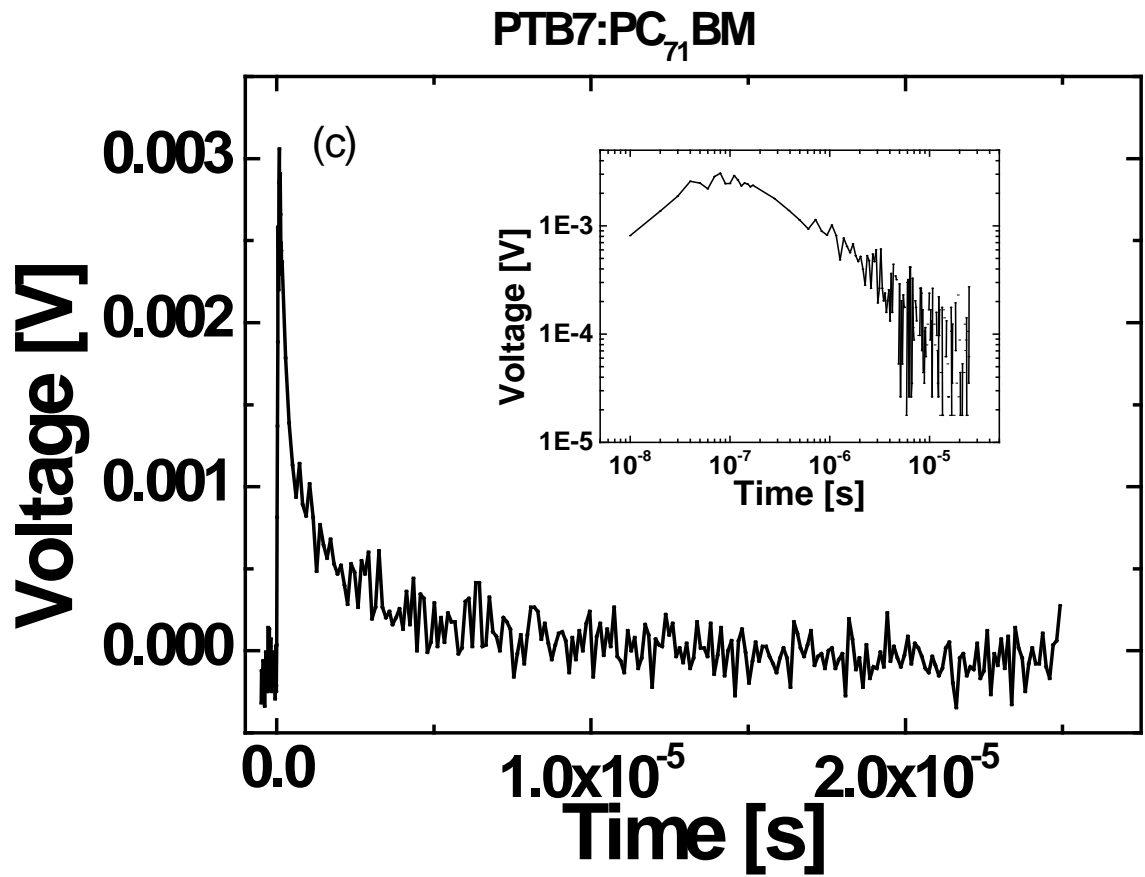

**Supplementary Figure 6.** Low light intensity time-of-flight photocurrent transients recorded at 50 ohm load resistance in thick (a, b) PCDTBT:$PC_{71}BM$ and (c) PTB7:$PC_{71}BM$ solar cells. Thick films were used to ensure surface carrier photogeneration. Insets show the same photocurrent transients replotted in log-log representation. The transit time kinks are not observed in any device. The decaying, featureless photocurrent is a typical signature of strongly dispersive transport.

**Film thickness dependence of carrier mobility**

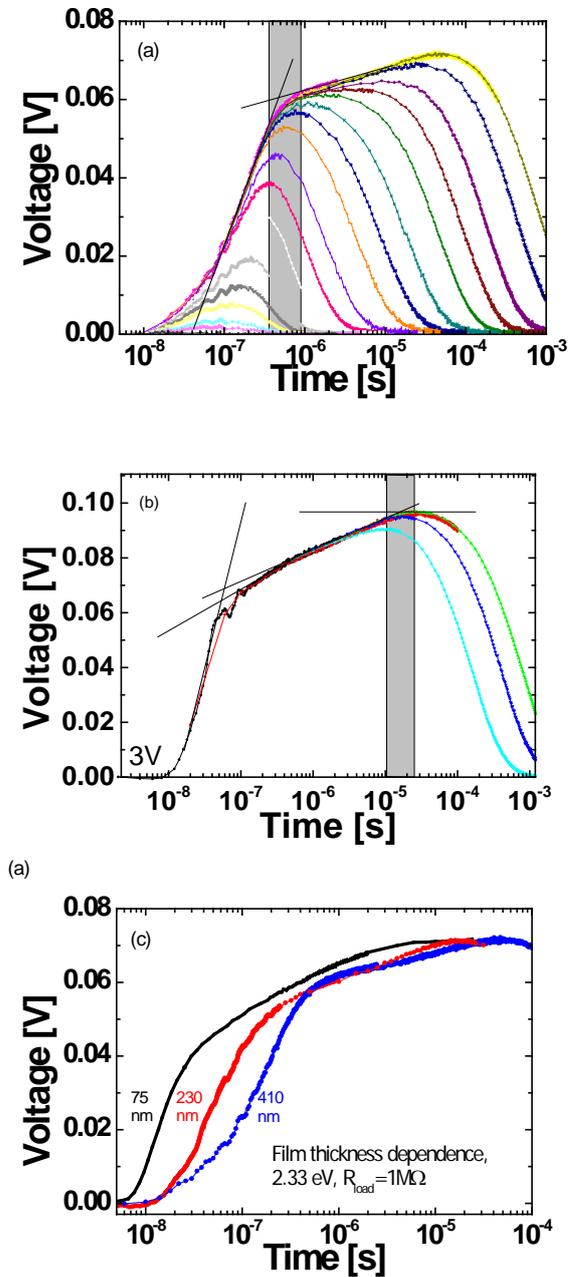

**Supplementary Figure 7.** RPV transients in PCDTBT:PC$_{71}$BM solar cells. (a) Typical RPV transients measured in a solar cell with a 410nm thick active layer. These transients are measured at built-in field and they do not saturate as a function of load resistance due to very long extraction times. In (b), the same devices were measured at 3V applied voltage. These transients show the voltage saturation at largest load resistance. (c) A comparison of RPV transients measured at large $R_{load}$ = 1 MΩ voltage in 75 nm, 230 nm and 410 nm thick active layers. From the transients the film thickness dependent mobility of holes and electrons is estimated and shown in the main text.

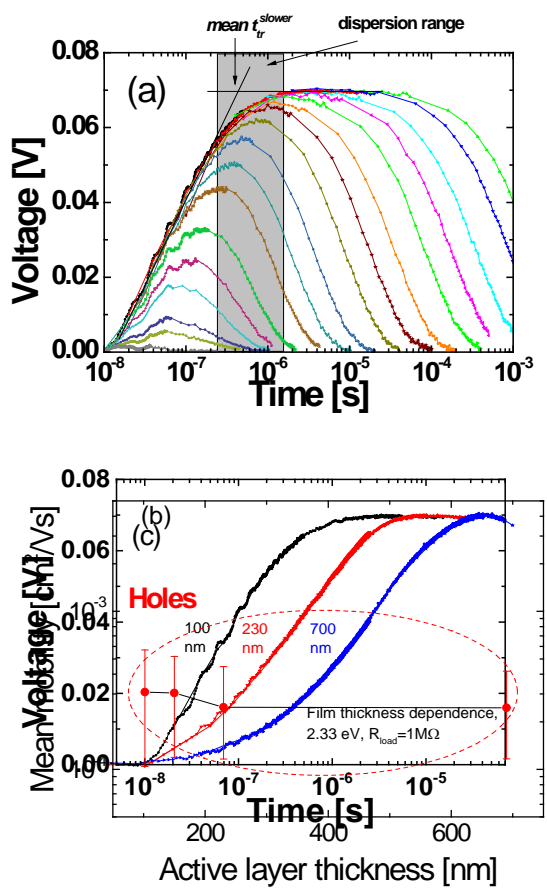

**Supplementary Figure 8.** RPV transients in PTB7:PC$_{71}$BM solar cells. (a) Typical RPV transients measured in a solar cell with a 100 nm thick active layer. (b) A comparison of RPV transients measured at large $R_{load}$ = 1 MΩ voltage in 100 nm, 230 nm and 700 nm thick active layers. (c) Active layer thickness independent hole mobility.

**Electric field dependence of carrier mobility**

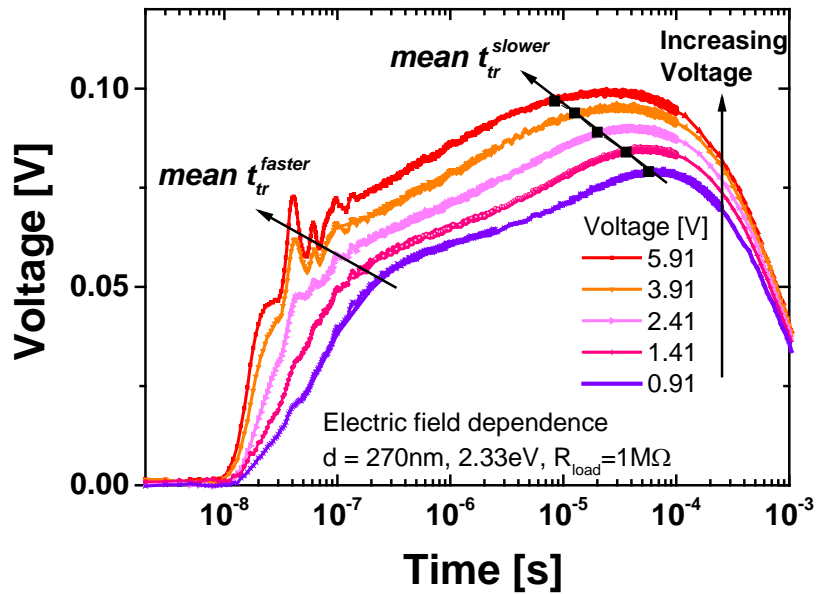

**Supplementary Figure 9.** Experimentally measured RPV transients at large $R_{load}$ = 1 MΩ at different applied voltages in the 270 nm thick PCDTBT:PC$_{71}$BM solar cell. The electric field dependent measurements in thin (75 nm) solar cells could not be obtained due to the electron mobility being too high with the transit times reaching the limiting time scale of the experimental setup. It demonstrates the shift of carrier transit shoulders to shorter timescales as the voltage is increased. From these transients the electric field dependent electron and hole mobilities are estimated and shown in the main text.

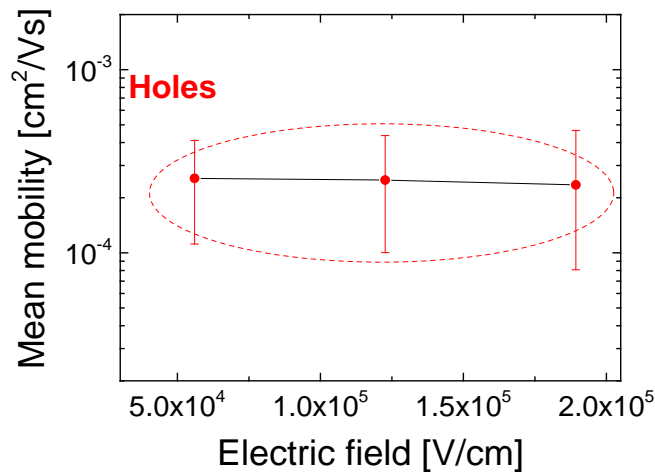

**Supplementary Figure 10.** Mobilities measured at varying electric fields on a PTB7:PC$_{71}$BM solar cell with active layer thickness of 150 nm. The mean mobility and dispersion range are independent of the electric field.

**Photon energy independent RPV transients**

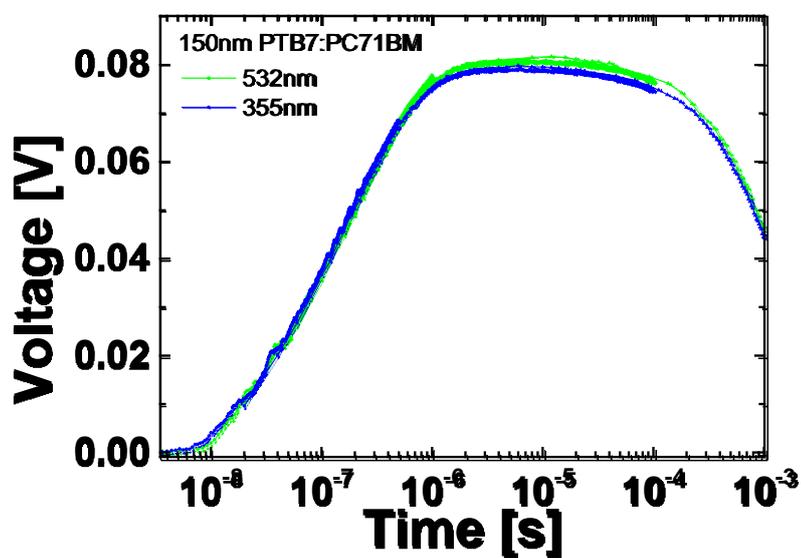

**Supplementary Figure 11.** Transients measured on a PTB7:PC$_{71}$BM solar cell at two different laser wavelengths. The transients are nearly identical, directly demonstrating that the charge carrier mobility is independent on photon energy and proves the absence of hot carrier contribution to charge extraction.

**RPV transients measured on PCBM-only diodes**

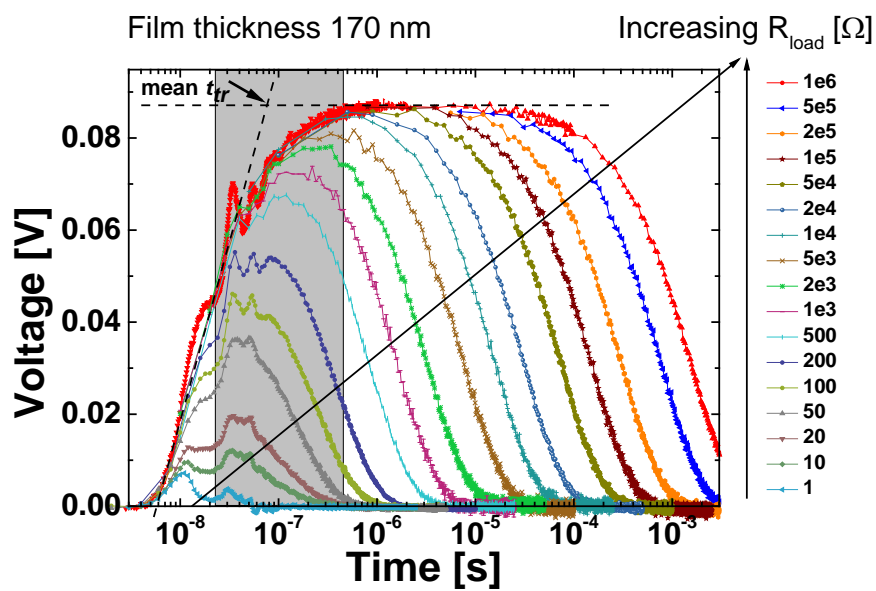

**Supplementary Figure 12.** Experimentally measured RPV transients in $PC_{71}BM$-only diodes. Only one shoulder is observed (at 77 ns, marked by an arrow) with a corresponding mean electron mobility of $5.5 \times 10^{-3}$ $cm^2$ $V^{-1}$ $s^{-1}$ with dispersion range from $1 \times 10^{-3}$ $cm^2$ $V^{-1}$ $s^{-1}$ to $2 \times 10^{-2}$ $cm^2$ $V^{-1}$ $s^{-1}$. The electron mobility is similar to the mobility of faster carriers in PCDTBT:$PC_{71}BM$ solar cells, identifying the faster carriers as electrons as would be expected.

**Probing deep trap states**

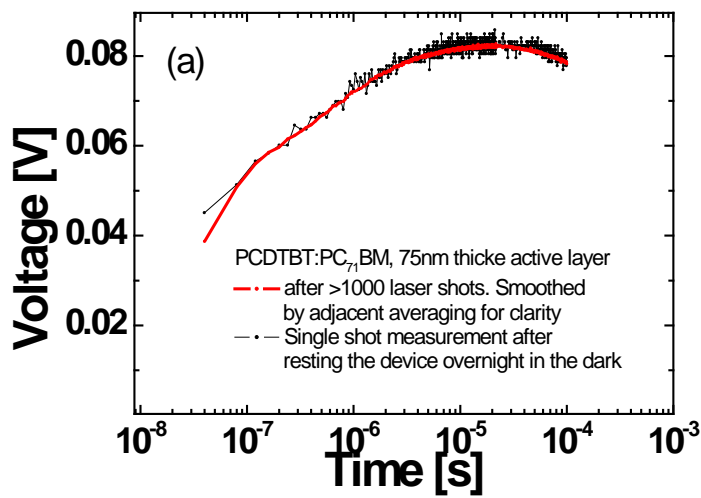

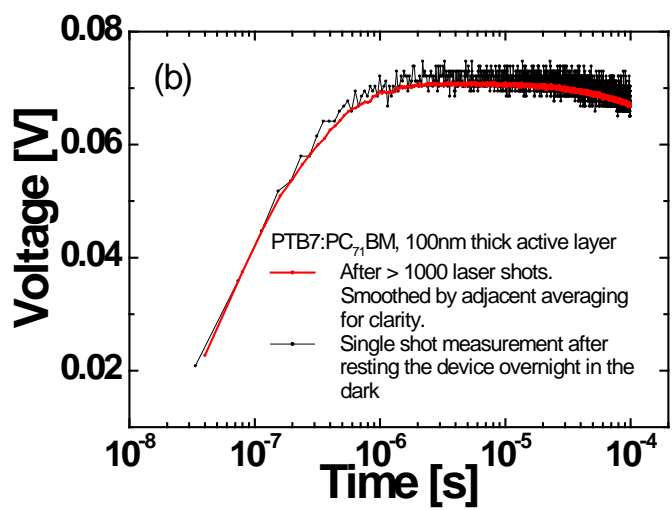

**Supplementary Figure 13.** Transients measured to detect deep trap states, or charging of the devices with repetitive laser shots. Deep traps cause RPV signals to be reduced in magnitude when the traps are filled in (main text, Figure 2b). Here, a cell is allowed to rest overnight, allowing any deeply trapped charges to escape. The change in voltage is negligible, demonstrating the absence of deep traps in this system.

**Non-dispersive transport simulations**

Non-dispersive simulations were performed using one dimensional continuity equations for electron and hole number densities. All quantities are scaled such that they are dimensionless, where dimensionless quantities are denoted with a prime.

Distances are scaled to the film thickness: $x' \equiv x/d$. Times are scaled to the transit time calculated for the fastest mobility: $t' \equiv t/t_{tr}$. The voltage scale is the internal voltage: $U' \equiv U/U_{internal}$, where $U_{internal}$ is the sum of the built in voltage and the applied voltage. This system of units requires that the normalized faster carrier mobility is $\mu'_{faster} = 1$.

Charge is normalized to the charge on the electrodes $CU$, while number density is normalized to $CU$ per volume: $n' \equiv enSd/CU$, where $S$ is the surface area of the device. Current is normalized to $CU$ per transit time: $i' \equiv it_{tr}/CU$. The circuit resistance is $R' \equiv RC/t_{tr}$.

The Einstein relation for diffusion gives a dimensionless temperature $T = kT/eU_{internal}$. The recombination coefficient is normalized to the Langevin rate: $\beta' \equiv \beta/\beta_L$.

In the above system of units, the equations become:

$$j'_p = \mu'_p E' p' - \mu_p' T' \frac{\partial p'}{\partial x'} \tag{1}$$

$$j'_n = \mu'_n E' n' + \mu_n' T' \frac{\partial n'}{\partial x'} \tag{2}$$

$$\frac{\partial p'}{\partial t'} + \frac{\partial j'_p}{\partial x'} = -\beta(\mu'_p + \mu'_n) n' p' \tag{3}$$

$$\frac{\partial n'}{\partial t'} - \frac{\partial j'_n}{\partial x'} = -\beta(\mu'_p + \mu'_n) n' p' \tag{4}$$

$$\frac{\partial^2 U'}{\partial (x')^2} = n' - p' \tag{5}$$

$$E' = -\frac{\partial U'}{\partial x'}. \tag{6}$$

The boundary conditions for Poisson's equation are:

$$U'(t, 0) = V' \tag{7}$$
$$U'(t, 1) = 0, \tag{8}$$

where $V'$ is the voltage across the semiconductor:

$$\frac{dV'}{dt'} = \frac{1 - V'}{R'} - j_c' \tag{9}$$

$$j'_c = \int_0^1 j'_p(x) + j'_n(x) \, dx. \tag{10}$$

We use a finite volume method, so the boundary conditions for the transport equations are expressed in terms of the fluxes $j'_p$ and $j_n'$ at each electrode (a total of four fluxes, for two types of carrier each at two boundaries). Since the experiment is conducted under reverse bias, we assume no injection is possible. This immediately sets two such edge fluxes to zero. The other two represent charge *extraction* and are described by the local drift current $j'_p = \mu'_p E' p'$ (and similarly for electrons).

The initial condition for the number density in the Beer-Lambert case is:

$$n'(0, x') = p'(0, x') = L' \alpha' e^{-\alpha' x'}, \tag{11}$$

with $Q'_{ph} = L'(a - e^{-\alpha'})$ and $\alpha' \equiv \alpha d$; or alternatively, by the condition of uniform generation

$$n'(0, x') = p'(0, x') = Q_{ph}. \tag{12}$$

The initial condition for voltage is $V' = 1$.

The above system of equations assumes a simple bimolecular recombination model. In the RPV experiment, the charge concentrations are deliberately kept low, so recombination events are rare simply because carriers are unlikely to meet. Therefore, the results are insensitive to changes in the recombination model. Typically, Langevin recombination was assumed, but we also tested non-Langevin recombination with reduction factors as low as $\beta/\beta_L = 10^{-3}$ and the results are unchanged. The only exception to this rule is when repeated laser shots add charge faster than it can be extracted, accumulating large charge concentration. That case is discussed in the main text.

The Einstein relation was used to calculate the diffusion coefficient at room temperature. The results are insensitive to changes in the diffusion coefficient because there are strong electric fields inside the device. The fields are strong because the low concentration of charge prevents field screening, and so the drift current dominates and the concentration of diffusion is insignificant.

**Dispersive transport simulations**

To simulate dispersive transport, the solver is modified as follows. Our objective was to reproduce the experimental signatures of dispersive transport (e.g. a spread of mobilities, and Scher-Montroll plots in time-of-flight) using a simple model with the fewest number of free parameters. We selected a multiple trapping model, whereby carriers are partitioned into a conduction band distribution $n(t,x)$ and a trapped distribution $n_t(t,x,E)$, where $E$ is the energy below the conduction band. The trapping and de-trapping terms in the continuity equation are:

$$\frac{\partial n}{\partial t} = \int_{-\infty}^{0} \frac{e^{E/k_B T} n_t(E)}{\tau_r} - \frac{n g(E)}{\tau_c} \, dE \quad (13)$$

$$\frac{\partial n_t(E)}{\partial t} = -\frac{e^{E/k_B T} n_t(E)}{\tau_r} + \frac{n g(E)}{\tau_c}, \quad (14)$$

where $k_B$ is the Boltzmann constant, $T$ is temperature, $\tau_r$ is a release time, $\tau_c$ is a capture time, and $g(E)$ is the density of states. We used an exponential density of states,

$$g(E) = \frac{e^{E/E_{trap}}}{E_{trap}}. \quad (15)$$

Three parameters are required to describe the trapping process: the release and capture times $\tau_r$ and $\tau_c$, and the width of the density of states, $E_{trap}$.

Numerically, these equations are discretized with respect to energy $E$ by introducing $N$ trap distributions partitioned as

$$n_t = \sum_{i=0}^{N-1} n_{t,i}, \quad (16)$$

and similarly for holes. Each trap distribution represents a particular energy level.

**Supplementary Table 1:** Simulation parameters for Figure 2 (a) in the main text.

| Parameter | Value |
| --- | --- |
| $E_{trap}/kT$ (for faster carriers) | 1.3 |
| $E_{trap}/kT$ (for slower carriers) | 1.5 |
| $\tau_c/t_{tr(faster)}$ (both carriers) | $10^{-4}$ |
| $\tau_r/t_{tr(faster)}$ (both carriers) | $10^{-6}$ |
| Beer-Lambert absorption $\alpha d$ | 2 |